# Magnetic oxides

*Daniel I. Khomskii, email: khomskii@ph2.uni-koeln.de, University of Cologne, Cologne, Germany*

*Sergey V. Streltsov, email: streltsov.s@gmail.com, M.N. Mikheev Institute of Metal Physics, Ekaterinburg, Russia*

**Key points.**

- There are different factors such as the spin-orbit coupling, intra-atomic Hund's exchange, crystal-field splitting and the Jahn-Teller effect controlling electronic and magnetic properties of isolated magnetic atoms.

- Intersite effects e.g. electron tunneling between ions, strong electronic correlations or features of the Fermi surface (in metals) define exchange interaction between magnetic ions.

- Exchange coupling can be anisotropic (different between different spin components). Competition between different exchange bonds can result in very unusual magnetic structures and can lead to different anomalous magnetic states such as spin liquids and spin ice.

- There are insulating magnetic oxides (magnetoelectrics and multiferroics), in which one can influence magnetic state by electric field and, vice versa, electric polarization can be controlled by magnetic field.

- Various transitions including structural and metal-insulator transitions are possible in magnetic oxides. Moreover, some of them can be driven into a superconducting state.

- Magnetic oxides have many applications, e.g. they can be used as magnetic field sensors, for nondestructive control and in spintronics.

**Abstract.** In this article we give a general survey of the main properties of magnetic oxides – mostly the oxides of transition metals, but sometime also containing rare earths ions. This is a very rich class of materials, among which there are insulators and metals, systems with insulator-metal transitions, and there are among them even high-temperature superconductors. One of the main features of these compounds, which attract to them special attention and which serve as a basis of many applications, are their rich magnetic properties. In this article we discuss the main physical effects determining their behaviour, and describe in detail especially their magnetic properties, but not only. After shortly discussing the basic structure of isolated magnetic ions, we concentrate on the collective effects depending on the interaction between sites, especially exchange interaction, giving rise to different magnetic properties: different types of magnetic ordering in conventional systems, but also more exotic states such as spin liquid states in frustrated systems. We also cover related phenomena in magnetic oxides, such as magnetoelectric and multiferroic behaviour, and discuss at the end their diverse useful properties serving as a basis of many applications.

## 1. Introduction

Magnetism is a big and important part of fundamental and applied physics.. There exist many different magnetic materials. There are among them many metals, alloys and intermetallic compounds, but also many insulators displaying rich and interesting magnetic properties. For many applications insulating magnetic materials are preferred, because they have less losses, can be used in combination with optics, etc.

Among insulating magnetic systems oxides represent probably the largest and most important group with rather diverse properties, and they already have many applications. The origin and magnetic properties are often rather different in metals and in insulators. Basic physics of magnetic oxides is quite rich, see e.g. the phase diagram of magnanites shown in Figure 1, and one can demonstrate on it many fundamental physical phenomena in solid state physics. Discussion of these the main topic of the present chapter.

When speaking of magnetic oxides, one typically has in mind oxides of transition metals (TM), most often of *3d* elements, but *4d* and *5d* and rare earths (RE) compounds become more and more popular and these systems have brought many interesting effects to the field. This chapter will be mainly focused on the most important case of TM oxides, but some comments are made also about RE compounds.

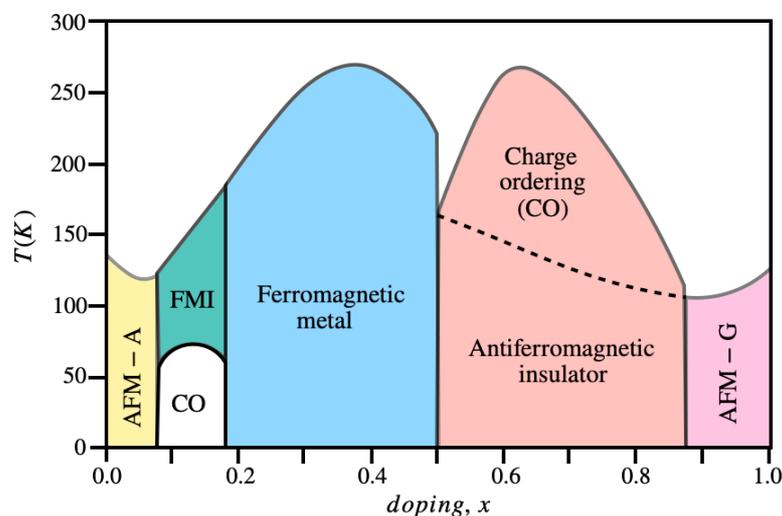

*Figure 1. Phase diagram of La$_{1-x}$Ca$_x$MnO$_3$. AFM-A and AFM-G stand for antiferromagnetic structure of the types A and G, see Fig10 (a); FMI stands for ferromagnetic insulator, CO – charge ordering.*

## 2. Isolated magnetic ions in a crystal

TM oxides are mostly insulating systems. The crucial role in their properties is played by electrons of partially-filled *d*-shells (or *4f*-shells for RE systems). The *d*-electrons, and even more so *f*-electrons, have relatively small radius, and they often can be treated as localized at corresponding ions. Such localized electrons have simultaneously also localized magnetic moments, and these are mainly responsible for magnetism of corresponding substances.

The *d*-electrons have an orbital moment $l = 2$, i.e. for an isolated atom or ion they are *2l+1*=5-fold degenerate, $l_z = -2, -1, 0, +1, +2$. In crystals such as TM oxides (NiO; Fe$_2$O$_3$ etc) TM ions are usually located in ligand "cages" formed by anions, like O$^{2-}$, F$^-$ etc., e.g. they sit in oxygen octahedra or tetrahedra. In this situation the point symmetry becomes lower: instead of spherical symmetry for isolated ions it becomes cubic or even lower due to the influence of ligands, such as oxygens. In cubic symmetry (regular octahedra or tetrahedra) the 5-fold degenerate *d*-levels are split into a doublet $e_g$ and triplet $t_{2g}$, doublet lying higher than triplet in octahedra (right part of Figure 2a); the order of these levels being reversed in tetrahedra, left part of Figure 2a. There exist also other coordinations, e.g. trigonal prisms or trigonal bipyramids, the crystal-field (CF) splitting in these cases is shown in Figure 2a.

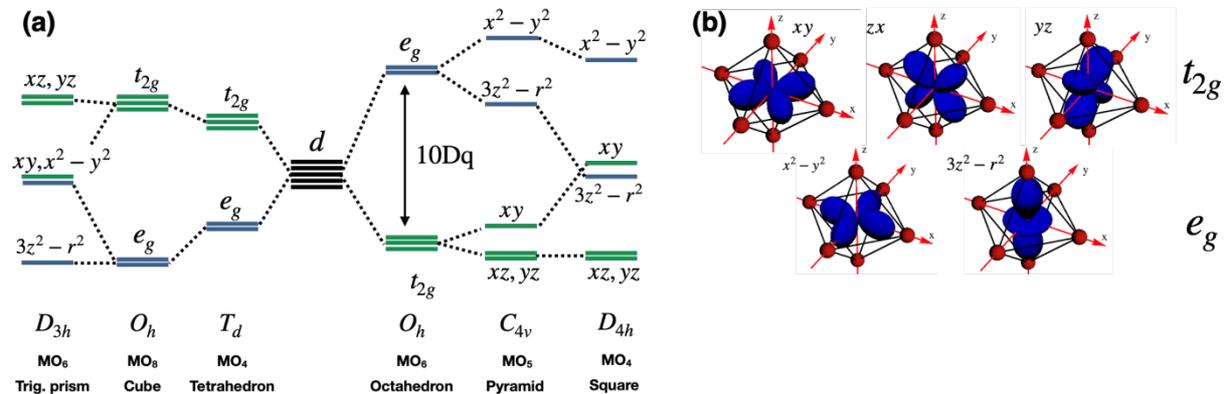

Figure 2. (a) Crystal-field splitting of d-orbitals for different types of local surrounding of TM ions. (b) Five d-orbitals, the situation of octahedral surrounding is shown. Ligands are red balls.

This CF splitting (for TM ions in octahedra or tetrahedra it is usually denoted as $\Delta_{CF} = 10Dq$) is caused by two mechanisms: (1) by the Coulomb interaction of respective electron density of corresponding orbitals. The typical shape of the five d-orbitals is shown in Figure 2b, with the charges of surrounding ligands (here $O^{2-}$ ions), and (2) by the covalency of d-electrons with the *2p* orbitals of oxygens. Both mechanisms usually lead to qualitatively the same CF splitting, the second mechanism (*d-p* covalency) typically playing more important role.

These CF levels are filled one by one by respective d-electrons. Then one has to take into account different interactions between *d*-electrons, the most important being here the Hund's rule interaction which can be written as, e.g., $-J_H S_{i\alpha} S_{i\beta}$ (*i* is the site index, α and β – orbital indices and α ≠ β, S is spin, $J_H$ is the Hund's intra-atomic exchange parameter), and also the spin-orbit coupling (SOC), which for single electron is $\zeta \vec{l}\vec{s}$, but which for the resulting many-electron terms with the total spin $\vec{S}$ and total orbital moment $\vec{L}$ per site is usually written as $\lambda \vec{L}\vec{S}$, where $\lambda = \zeta/2S$ is the spin-orbit coupling constant [Landau and Lifshitz, 1965]. SOC strongly increases in going down in periodic table, $\lambda \sim Z^2$ [Landau and Lifshitz, 1965], where Z is the atomic number of the element. It is relatively weak for *3d* ions ($\lambda \sim 20 - 70$ meV) [Abragam and Bleaney, 1970], but it becomes much stronger, up to ~0.4 eV, for *4d* and *5d* ions; it often determines the properties of these compounds.

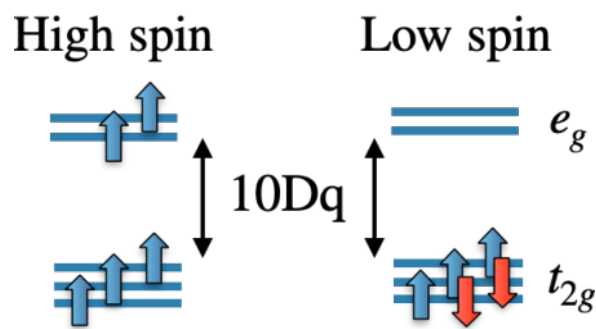

Figure 3. Two different spin states in case of $d^5$ electronic configuration.

The usual filling of one-electron levels for weak SOC is determined by the interplay of CF splitting and the Hund's interaction (10Dq vs $J_H$). If $J_H > 10Dq$, which is mostly the case of *3d* oxides, electrons of TM ions with $d^n$ configuration fill *d*-levels one by one from below such as to give the maximum total spin (e.g. $S=5/2$ for $d^5$ configurations, Figure 3 left part. This is the so called high-spin (HS) state, typical for $3d$ ions. But if $J_H < 10Dq$, which is usually the case for $4d$ and $5d$ ions, one would better fill the lowest CF levels, which e.g. for the $d^5$ configuration would give not $S=5/2$ as for HS configuration left part of Figure 3, but $S=1/2$, right part of Figure 3 (for other electron configurations the exact criterion could be modified, e.g. 10Dq vs $2J_H$, or vs $3J_H$). This is e.g., the situation for $Ru^{3+}$ in $RuCl_3$ and $Ir^{4+}$ in $Na_2IrO_3$ – the very interesting materials with the strongly bond-dependent (so called Kitaev) exchange interaction and with nontrivial properties, see Section 4. This is the low-spin (LS) state, typical for $4d$ and $5d$ oxides, but also met in some $3d$ systems, e.g. those with the $d^6$ configuration ($Fe^{2+}$, $Co^{3+}$) in octahedra, for which six $d$-electrons may completely fill the lowest $t_{2g}$ levels making the LS state of these ions nonmagnetic, $S=0$. This is exactly the situation in the famous $LaCoO_3$, which is non-magnetic at low temperature, but experiences the spin-state transition to magnetic state under heating (particular scenarios are still debated), due to a thermal expansion resulting in decrease of the CF splitting.

The spin-state transitions (known also as spin crossovers in chemistry) are very important for geoscience, since under the pressure (deep in the Earth) TM-ligand distances are strongly decreased and therefore the CF splitting in contrast is increased, transforming TM ions into the LS state. Such transitions are accompanied by the change of the volume and manifests itself by variation of the density affecting seismic compressional velocities.

In certain situations the electronic filling of CF levels in symmetric situation (e.g. regular octahedra) may be such that there remains an orbital degeneracy, either double (one or three electrons on the $e_g$ levels) or triple (partially-filled $t_{2g}$ levels). In this case, according to the famous Jahn-Teller (JT) theorem, the symmetry of the system would spontaneously lower, leading to splitting of degenerate levels and giving a gain in electronic energy, linear in distortion, overcoming the quadratic loss of elastic energy [Kaplan and Vekhter 1995]. The spin-orbit coupling can affect the Jahn-Teller effect [Streltsov and Khomskii 2020]. The well-known ions with strong Jahn-Teller effect are e.g. $Mn^{3+}$ and $Cu^{2+}$, important in manganites with colossal magnetoresistance (CMR) and in High-Tc superconductors, see below.

## 3. Intersite effects

While the ground state can be defined by the single-site physics, the intersite effects of different types are extremely important – in particular this is the intersite electron hopping. The actual hopping could lead to conductivity, eventually making the system metallic. The virtual hopping due to a quantum tunnelling in insulating situation can give rise to exchange interaction. In this case, when we allow for the change of the number of electrons at a site, the intra-atomic Coulomb interaction between electrons starts to play crucial role. In the simplest form, first ignoring orbital degrees of freedom, one can describe the situation by the Hubbard model

$$H = -\sum_{ij\sigma} t_{ij} c_{i\sigma}^\dagger c_{j\sigma} + U \sum_i n_{i\uparrow} n_{i\downarrow} \qquad (1)$$

where $i,j$ numerate lattice sites and $\sigma$ is the spin index. The first term describes the intersite hopping of (here nondegenerate) electrons, and the second term is the on-site Coulomb (Hubbard) interaction. For one electron per site we can have here two situations. If $U \ll t$, the dominant role is played by the first term in (1), intersite hopping, which after Fourier transform gives the partially-filled (here half-filled) energy band and consequently a metallic state with itinerant electrons. Interaction $U$, assumed here to be weak, can then be treated e.g. by perturbation theory. This is the standard band theory of solids.

In the opposite case of $U \gg t$, however, the situation is very different. To avoid strong Coulomb repulsion the electrons would remain localized one per site. This is the state apparently first suggested by Peierls, the theory of which was later developed largely by Mott; this state is now called Mott insulator [Khomskii, 2014]. It is this state which is the basic state of most magnetic oxides: the presence in it of localized electrons also means the presence of localized spins, or localized magnetic moments. Very schematically one can describe the general situation on a simplified diagram showing the evolution of one-electron bands with increasing Coulomb (Hubbard) repulsion $U$, characterising on-site electron-electron repulsion. Here the half-filled metallic band existing for $U = 0$ transforms into two bands: the lower and upper Hubbard bands for $U$ more than the critical value $U_c$ (of order of the bandwidth $W = 2zt$, where $z$ is number of nearest neighbours). These bands are divided by the energy gap $\sim U$, so that the lower band is filled and the upper one is empty. (One has only to remember that it is an effective picture, with the gap created by electron-electron repulsion, in which case the simple one-electron description is, strictly speaking, not valid. Specifically, the "volume", the number of states in each Hubbard band is not a constant but depends on the total number of electrons).

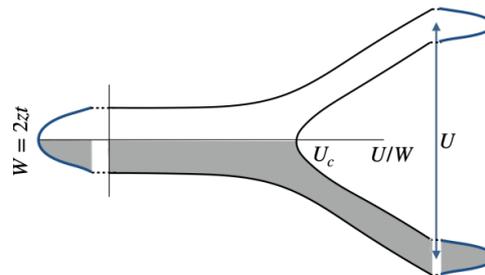

*Figure 4 Sketch illustrating formation of the lower and upper Hubbard bands due to the on-site electron-electron repulsion (characterized by the Hubbard's term U). The schematic bands and the density of states are shown.*

In realistic situations we have to take into account also the factors described in Section 2, mostly connected with orbital degrees of freedom of *d*-electrons. And of course, a lot depends on the type of respective crystal lattice (rock salt in monooxides like MnO, NiO; corundum structure $M_2O_3$, e.g. $Fe_2O_3$, $Cr_2O_3$; perovskites $AMO_3$, e.g. $LaMnO_3$; spinels with the general formula $AM_2O_4$, etc) [Imada et al., 1998].

Very important is also to realise that usually there are anions, such as oxygen ions, between TM ions in real TM oxides. Sometimes it is sufficient to describe the situation keeping only *d*-electrons, as we did above, but often *p*-electrons of oxygen play also an important and, in some cases, crucial role. For example, in perovskites the oxygen ions are located just in between TM ions, so that the effective electron hopping between TM ions occurs via *p*-states of intermediate oxygens. One can describe this situation by the effective *d-p* model, including also *p*-electrons of oxygen:

$$H_{dp} = \sum_{i\sigma} \varepsilon_d d^\dagger_{i\sigma} d_{i\sigma} + \sum_{j\sigma} \varepsilon_p p^\dagger_{j\sigma} p_{j\sigma} + \sum_{ij\sigma} t^{pd}_{ij} d^\dagger_{i\sigma} p_{j\sigma} + \text{h. c.} + U \sum_i n^d_{i\uparrow} n^d_{i\downarrow}, \qquad (2)$$

where $t^{pd}$ is the hopping between TM $d$- and ligand $p$-orbitals $\varepsilon_d$ and $\varepsilon_p$ are their energies. If necessary, one should include in this description also other interactions: between $d$- and $p$-electrons, $p$-$p$ interaction etc. If the oxygen $p$-levels are deep below $d$-levels of TM ions, one can exclude $p$-electrons and return to the description of the type of Hubbard model $\tilde{t}_{dd} \sim t_{pd}^2/(\varepsilon_d - \varepsilon_p)$, with the effective hopping $\tilde{t}_{dd}$ (more accurately one has to include here in the denominator also the effects of electron-electron interactions, which, however, may be not so important for large $\varepsilon_d - \varepsilon_p$). However if $p$-levels lie not far from the $d$-levels, the situation may change drastically. In this case the ground state may still be the same – electrons localised at TM ions, a magnetic insulator. But the lowest excited states are not those following from the simple Hubbard model (1) and shown in Figure 4, i.e. not the transfer of a $d$-electron from one TM ion to the other, the process $d^n d^n \to d^{n+1} d^{n-1}$, which costs the energy $U$, but the lowest charge-carrying excitations would be the transfer of a $p$-electron from the $O^{2-}$ ion to the $d$-shell of TM, the process $d^n p^6 \to d^{n+1} p^5$, with the creation of an oxygen hole. The energy cost of this process is called the change transfer (CT) energy [Sawatzky and Green, 2016]

$$\Delta_{CT} = E(d^{n+1}p^5) - E(d^n p^6). \tag{3}$$

Crudely speaking, this CT energy is, according to (2), $\Delta_{CT} = \varepsilon_d - \varepsilon_p + U$, but it can also depend on the $pd$ and $pp$-interactions. Correspondingly all correlated insulators can be subdivided into two big groups, see the famous Zaanen-Sawatzky-Allen diagram in Figure 5. Thus depending on the ration between the Hubbard and CT energies we can have two types of insulators: the usual Mott insulators if $\Delta_{CT} > U$, and CT insulators for small CT energy if $\Delta_{CT} < U$. As always, the insulating states exists if both $U$ and $\Delta_{CT}$ are larger than the electron hopping $t$. In the opposite case we can have metallic state for Mott part of the phase diagram, and rather nontrivial situation – possibly metallic but also sometimes insulating states, if we move on this phase diagram to the left, to the region of small or negative CT energy. We will discuss this situation below in Sections 8, 9.

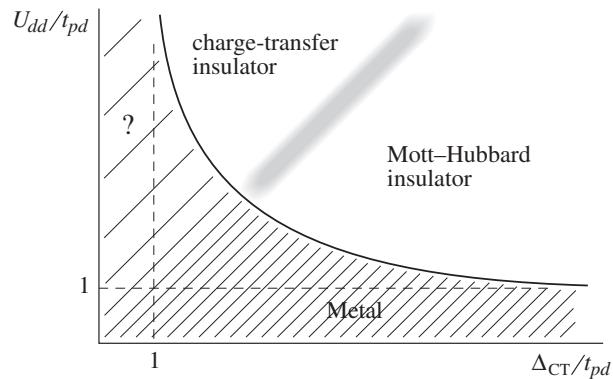

Figure 5. Zaanen-Sawatzky-Allen diagram. It classifies insulators according to the character of the states forming valence and conduction bands. Here $U_{dd}$ is the Hubbard on-site electron repulsion parameter, $\Delta_{CT}$ – charge-transfer energy, $t_{pd}$ – tunnelling amplitude (hopping) from p-orbital of a ligand to d-orbital of TM ion.

## 4. Exchange interaction

There are various mechanisms which contribute to exchange interaction between magnetic centers in oxides. The simplest Hamiltonian to describe this interaction was proposed by Werner Heisenberg,

$$H = \sum_{i>j} J_{ij}\vec{S}_i\vec{S}_j \qquad (3)$$

who suggested that the non-local part of the Coulomb repulsion between electrons at different sites (numerated by indexes $i$ and $j$) is responsible for this interaction; the strength of which is controlled by exchange constants $J_{ij}$. However, it was later understood that in most cases the mechanism of exchange interaction in magnetic insulators such as TM oxides is very different. In these systems electrons are mostly localized at their sites, but due to the virtual hopping, described by the first term in (1) they can tunnel between magnetic ions, which leads to decrease of total energy. However, the Fermi statistics and intra-atomic Hund's exchange interaction strongly affect this tunnelling. In elemental example sketched in Figure 6a the hopping between FM ordered spin is forbidden by the Pauli principle and for this state we do not gain energy; therefore quantum tunnelling will stabilize AFM interaction with $J \sim 4t^2/U$ [Goodenough, 1963]. If the hopping occurs directly between $d$-orbitals one speaks about the direct exchange, but often magnetic ions are too far away from each other and $d$-electrons can hop only via ligand's $p$ states. This is what is called the superexchange interaction. In real materials the situation is much more complicated, there are in fact many orbitals (five in case of TM) and there can be quite a few possible hopping paths resulting in various contributions to the total exchange, see also Section 5, but still the strongest exchange (for appropriate filling of orbitals) is AFM as in Figure 6a. This is why most of insulating TM oxides are AFM, and there are only few FM ones among them, such as, e.g., YTiO$_3$ or NaCrGe$_2$O$_6$, with typically low Curie temperatures [Streltsov and Khomskii, 2017].

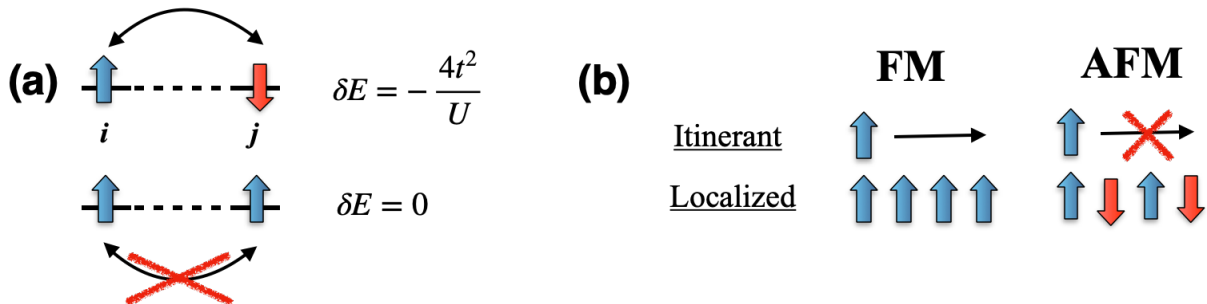

Figure 6. (a) Sketch illustrating mechanism of the direct exchange or superexchange (for one active orbital per site): in case of FM order tunnelling processes are forbidden and cannot lower the total energy of the system, but this energy gain exist for antiparallel spins, i.e. this process leads to AFM coupling. (b) Mechanism of the double exchange leading to FM: itinerant electrons easily propagate through the lattice in case of FM order, thus gaining kinetic (band) energy, since they follow the first Hund's rule at every site.

Going to metallic magnetic oxides one needs to mention two other important exchange mechanisms. The first one is the so-called double exchange. Suppose there are two groups of electrons - localized and itinerant, as, e.g., in Sr doped LaMnO$_3$ where $t_{2g}$ electrons are localized at Mn sites, while $e_g$ orbitals are very different – these $d$-orbitals are directed exactly to each other in perovskite structure of LaMnO$_3$ and therefore corresponding bands are much wider. Moreover, due to not complete filling these $e_g$ bands are metallic. Thus, there are localized $t_{2g}$ electrons forming localized magnetic moments, and $e_g$ ones moving on the background of these moments as illustrated in Figure 6b. It is clear that it is much easier for them to propagate if all moments are FM, otherwise at every other Mn site $e_g$ electrons appear in the "anti-Hund" configuration. Thus, there is a FM contribution to the exchange interaction due to the presence of such metallic electrons. Again, in real materials there will be competition between the FM double exchange and (typically) AFM superexchange, and the outcome, i.e. the resulting sign of the interaction, is not obvious. In particular, this can result in canting of the spins or in the formation of helicoidal magnetic structures.

The last exchange mechanism to mention is the Ruderman-Kittel-Kasuya-Yosida (RKKY) interaction which is characteristic for metallic systems such as metallic RE compounds. A localized moment is known to generate in metals an oscillating spin polarization of conduction electrons. Another magnetic moment interacts with these electrons and therefore feels the presence of the first moment. The resulting RKKY exchange is is slowly decaying with distance and is oscillating in space, so that the sign of the RKKY exchange can be both FM or AFM [Yosida 2010].

It has to be noted that the equation (3) often turns out to be oversimplified, there can appear higher order terms in spin variables, but also $J$ is allowed to be not a scalar, but a tensor $J^{\alpha\beta}$, where $\{\alpha, \beta\} = \{x, y, z\}$. Then the diagonal part equal for all spin components gives isotropic exchange $J = J^{xx} = J^{yy} = J^{zz}$, while off-diagonal ones give anisotropic interactions, for example generate Dzyaloshinskii-Moriya (DM) interaction, which can be rewritten in a compact form as $H_{ij}^{(DM)} = \vec{D}_{ij} \cdot (\vec{S}_i \times \vec{S}_j)$, from which it becomes clear that the DM interaction favors spin canting. This is an extremely important interaction, which is responsible for appearance of various non-collinear magnetic structures, in particular skyrmions, and it can result in formation of spontaneous electric polarization in multiferroics via the so-called inverse DM effect, see Section 7.

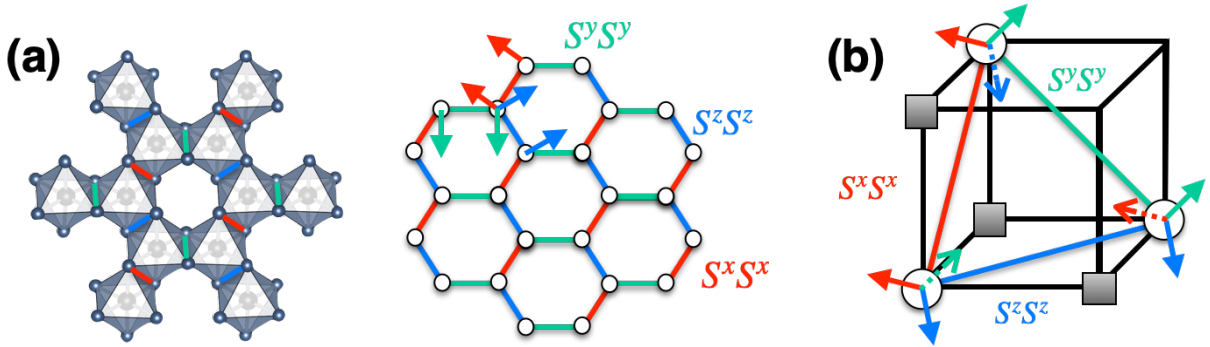

*Figure 7. Left: the crystal structure of one of the Kitaev material, Na$_2$IrO$_3$: Ir is in O$_6$ octahedra, which share edges shown by different colors. Right: corresponding spin lattice with the bond depended exchange interaction: for green bond the largest coupling is between S$^y$ components, for red – S$^x$, and for blue - S$^z$. (b) Fragment of face-centered cubic lattice, which form magnetic metals (circles) in double perovskite structure, e.g. in Ba$_2$CeIrO$_6$. In both situations strong frustration is observed.*

The microscopic origin of DM interaction is the SOC mentioned in Section 2, it couples the spin space and the real space and can result in anisotropy of the exchange interaction [Abragam and Bleaney, 1970]. This can not only generate off-diagonal, but can also make diagonal components of the exchange tensor very different. There can also exist a single-site anisotropy, which result in a situation, when instead of vectors in (3) one has, e.g., only $z$ components of spins in effective spin Hamiltonian, i.e. $H = \sum_{i>j} J_{ij} S_i^z S_j^z$. This is the famous Ising model. Another more exotic situation is realized in Kitaev materials with a layered honeycomb structure, where due to strong SOC, specific lattice geometry and particular $t_{2g}^5$ electronic configuration the exchange interaction along some of the TM-TM bonds is large for $z$ components of spins, while along other bonds it is large for $x$ or $y$ components, see Figure 7a. In fact strong Ising-like anisotropy and bond-depended interaction were found in many oxides, e.g. Na$_2$IrO$_3$, Li$_2$IrO$_3$ [Takagi et al., 2019]. These are typically systems with heavy *4d* and *5d*

TM ions (since the SOC is large for them), with various orbital fillings and very different lattice geometry, such as e.g. double perovskites shown in Figure 7b.

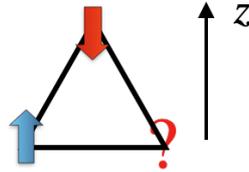

Figure 8. Illustration of frustration. If the exchange interaction is described by the Ising model and all spins are coupled antiferromagnetically, then they cannot satisfy all exchange bonds.

## 5. Frustrated magnetism, spin liquids, spin ice

The anisotropy of exchange interaction strongly restricts possible types of the ground states, and when it is combined with a lattice, which also poses its own constrains on the magnetic ordering, novel interesting physics may appear. A typical example is a triangle of AFM coupled spins, described by Ising model, shown in Figure 8. One readily realizes that it is impossible to satisfy all exchange bonds in this situation. This is what is called frustration: a system does not know how to order the spins. This results to a large degeneracy of the ground state. Already a triangular cluster has 6-fold degenerate ground state, but when we have a lattice of triangles this number rapidly increases and gives finite entropy at $T=0$. This has important implications for thermodynamics. There are many such lattices having a triangular motif: such are kagomé (e.g. $SrCr_9Ga_3O_{19}$), hyper-kagomé (e.g. $Na_4Ir_3O_8$), garnets (like $Gd_3Ca_5O_{12}$) or pyrochlore ($Pr_2Ir_2O_7$). But frustrations are also possible in other situations, e.g. AFM next-nearest neighbour exchange will frustrate AFM Ising chain, or bond-dependent anisotropic interaction in Kitaev model shown in Figure 7a leads to the situation when each spin "does not know what to do" – exchange interactions along each bond tend to direct it in its own way (along $x$, $y$, or $z$ direction).

Frustration suppresses a long-range magnetic order. One of its hallmarks often used in experimental physics is a large frustration parameter $f = |\theta_{CW}|/T_N$, where $\theta_{CW}$ is the Curie-Weiss temperature characterizing an average strength of exchange interaction, and $T_N$ is the Néel temperature (or spin glass transition temperature if a system finally freezes in a disordered state), i.e. the temperature at which a long-range magnetic order sets in. Frustration suppresses such a long-range order, and the parameter $f$ is a measure of this suppression. Typically, if $f \sim 5-10$, it might be taken as a signature of one might already expect to have some sort of frustration; stronger frustrations give (much) larger values of this parameter. Thus e.g. in $CuAl_2O_4$ $f \sim 65$, while in $SrCr_9Ga_3O_{19}$ it exceeds 150. (Note however that in any low-dimensional system the ordering temperature is strongly (logarithmically) suppressed by a weak interlayer of interchain exchange; this should not be mixed with the effect of frustrations).

Frustration can suppress a long-range magnetic order, while spins are still strongly coupled. By analogy with classical thermodynamics such a state with a long-range quantum mechanical entanglement of spins, but absence of a magnetic order is called a spin liquid (do not mix this state with a conventional paramagnetism, where disorder is induced by the temperature). The spin correlator can have a power-law dependence on distance $\langle \vec{S}_i \vec{S}_j \rangle \sim 1/|\vec{r}_i - \vec{r}_j|$ as in so-called algebraic spin liquids or decay exponentially $\langle \vec{S}_i \vec{S}_j \rangle \sim exp(-|\vec{r}_i - \vec{r}_j|/\xi)$ as e.g. in the Shastry-Sutherland model (physical realization of which is $SrCu_2(BO_3)_2$). Both theoretical and experimental studies of spin liquids in different models and compounds is one of the most actively developing directions in the modern condensed matter physics [Savary and Balents, 2017], each year hundreds of materials are claimed to host this mysterious state, but most of them finally reveal some sort of long-range magnetic order or appears to be atomically disordered. One of very promising system for realization of the spin-liquid is the mineral herbertsmithite ZnCu$_3$(OH)$_6$Cl$_2$ with the kagomé lattice of Cu$^{2+}$ ions wth $S=1/2$.

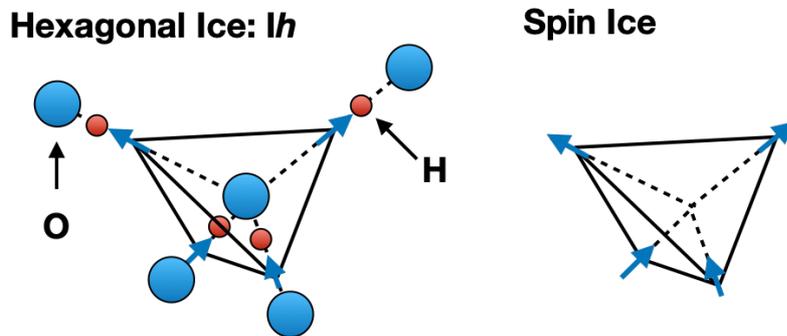

Figure 9. Conventional and spin ices. Arrows show displacement of hydrogens (or electric dipole moments) in real ice and spin directions in spin ice.

One more very interesting subject to mention in context of frustration is the spin ice systems. These are mainly pyrochlores like Dy$_2$Ti$_2$O$_7$. Tb$_2$Ti$_2$O$_7$, in which magnetic ions (Dy, Tb) form the lattice of corner-sharing tetrahedra (again with triangular motif) and are very strong Ising ions, with easy axes directed for each ion to the center of respective tetrahedron (or away from it) [Gardner et al., 2010]. This leads for certain interactions (including dipole-dipole interaction) to very strong frustration and to disordered ground state. This situation reminds the crystal structure of conventional ice, where oxygens form distorted tetrahedra, and hydrogen of H$_2$O molecules can be attached to one or to the other oxygen, i.e. can move to the center of oxygen tetrahedra or away from it. Spins in spin ice materials are directed in the same way as the electric dipoles formed by H$^+$ and O$^{2-}$ ions, see Figure 9; that is why the very term "spin ice". Many features of spin ice systems are quite exotic, in particular excitations in these systems have properties of magnetic monopoles.

## 6. Various orderings: magnetic, charge, orbital etc.

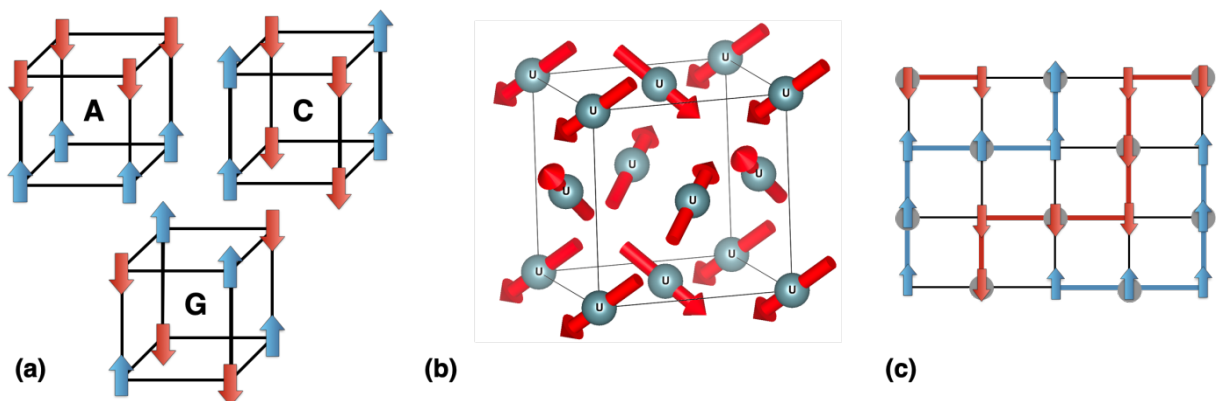

Figure 10. (a) typical types of AFM structures for a simple cubic lattice, (b) very complicated AFM structure of UO$_2$, (c) CE-type of AFM with FM zigzags, realised e.g. in half-doped manganites like La$_{0.5}$Ca$_{0.5}$MnO$_3$, cf. Figure 1

There can be very different magnetic structures in magnetic oxides. Many perovskites adopt simple two-sublattice structures as shown in Figure 10a (e.g. LaTiO$_3$ is AFM of G-type;

CaCrO$_3$ is C-type; LaMnO$_3$ is A-type), but there are also examples of very complex and seemingly counterintuitive structures, where for example spins on different sites can be nearly orthogonal to each other as in UO$_2$ (Figure 10b) or Bi$_2$Fe$_4$O$_9$. Often this is the result of competition between different exchange couplings, e.g. nearest and next nearest neighbours, but various anisotropic contributions to the exchange tensor $J^{\alpha\beta}$ can also strongly change the ground state magnetic structure; this factor is especially important for *4d* and *5d* oxides.

Spin is just one of the degrees of freedom in magnetic oxides. There are also other, such as orbital and charge degrees of freedom. If metallic ions have different oxidation states, e.g. there can be Mn$^{3+}$ and Mn$^{4+}$ ions in the same material, they can be ordered, in which case we speak of the charge ordering or more generally about formation of the charge density wave. There are many examples of magnetic oxides with charge ordering – e.g. Fe$_3$O$_4$, AlV$_2$O$_4$, Y$_2$Ni$_2$O$_6$ – and there can be very different mechanisms lying behind this effect. It can be connected with the nesting of the Fermi surface for initially metallic systems. The mechanisms of such charge ordering may be electron-phonon interaction or Coulomb interaction. There may also exist strong coupling of CO with other degrees of freedom. For example, in manganites such as PrBaMn$_2$O$_6$ and NaMn$_7$O$_{12}$ the AFM ordering, the so-called CE-type ordering (Figure 10c) coexists (and may be caused by) the checkerboard CO of Mn$^{3+}$ (grey circles in Figure 10c) and Mn$^{4+}$. The additional electron in $e_g$ shell of Mn$^{3+}$ ions provide strong FM coupling with two of its neighbours by the double exchange mechanism, with this electron occupying $3x^2 - r^2$ or $3y^2 - r^2$ orbitals for horizontal or vertical legs of the zigzag. These three ferromagnetically coupled spins are the example of trimer clusters ("trimerons") also invoked to explain the details of charge ordering in magnetite Fe$_3$O$_4$.

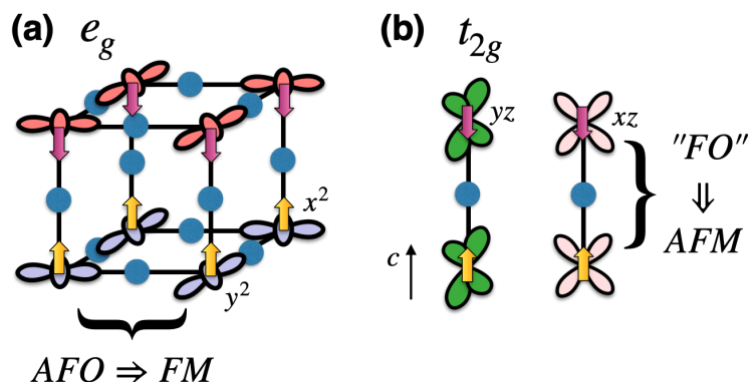

Figure 11. Orbital ordering in LaMnO$_3$ (Mn$^{3+}$, $t_{2g}^3 e_g^1$). Single $e_g$ electron occupies alternating $3x^2$-$r^2$ and $3y^2$-$r^2$ orbitals (dubbed as $x^2$ and $y^2$), which is the antiferro-orbital (AFO) ordering; it gives ferromagnetism in the ab-plane, see (a). Magnetic coupling along c is predominantly governed by the $t_{2g}$ orbitals. There is a strong overlap between the same $t_{2g}$ orbitals (via ligand's p-orbitals), which results in antiferromagnetism in the c-direction.

There exit yet another very important type of ordering in oxides - orbital ordering. It usually occurs in systems with orbital degeneracy and with strong JT effect, see Section 2. Local distortions due to the JT effect on different centers couple to each other, leading to the cooperative effect – cooperative JT ordering, or orbital ordering. This usually occurs at a special transition, which is simultaneously a structural transition with the reduction of the symmetry of a crystal. Besides its own features such orbital ordering also very strongly influences magnetic ordering. As was explained in Section 4, exchange interaction between the same half-filled orbitals (ferro-orbital situation) results in AFM, see Figure 6a, while it can be shown that if electrons occupy different orbitals (antiferro-orbital) and there is no overlap between these half-filled orbitals, then the exchange interaction can be ferromagnetic. One of the examples is LaMnO$_3$ with four *3d* electrons, one at $e_g$ and three at $t_{2g}$ orbitals. Antiferro-orbital ordering of one of $e_g$ orbital results in FM in the *ab* plane, while $t_{2g}$ states give AFM along the *c* direction, see Figure 11.

There are many examples of magnetic oxides with orbital ordering in addition to mentioned above manganate, but the most important is that in this class of materials – magnetic oxides – spin, orbital, charge and lattice degrees of freedom are often turn out to be interrelated and may affect each other. This feature is interesting from theoretical point of view and opens up new perspectives for technological application of such systems, for example one can change electronic properties by stress or magnetic properties by external electric field or light.

## 7. Magnetoelectrics and Multiferroics

In this article we mainly concentrate on magnetic properties of oxides. However, in several specific situations there appears quite nontrivial electric behaviour of these systems, closely connected with their magnetism, so that one can in principle influence magnetic state of the system by electric field. Such are the systems with (linear) magnetoelectric effect (ME), first proposed by Dzyaloshinskii in 1959 on symmetry grounds, the microscopic theory of which later on being elaborated by Moriya. This ME effect is based on the existence in systems with some special symmetries of the term in free energy of the type $\alpha_{ij}E_iH_j$ or $\tilde{\alpha}_{ij}P_iM_j$, where $P$ is an electric polarization and $M$ – magnetic moment, $E$ is electric and $H$ - magnetic field (we assume summating over repeated indices $\{i,j\} = \{x,y,z\}$). The presence of this term in the free energy leads to the possibility to induce magnetization

$$M_i = \alpha_{ij}E_j$$

by electric field or. vice versa, of electric polarization by magnetic field,

$$P_i = \alpha_{ij}H_j$$

The magnetoelectric tensor $\alpha_{ij}$ can have both symmetric and antisymmetric components; the symmetric one leads e.g. to the magnetic moment parallel to the applied electric field, and antisymmetric ME tensor – to the moment $M$ perpendicular to $E$.

Already a year after this prediction by Dzyaloshinskii the ME effect was found by Atsrov in Cr$_2$O$_3$, and later on it was actively studied in many materials, mostly in antiferromagnetic TM oxides. But the coupling of electric and magnetic properties is not restricted to the ME systems, in which one needs external fields to observe this effect. It turned out that there also exist systems in which magnetic and electric orderings, in particular ferroelectricity (FE), coexist without external fields. Such materials were called multiferroics (MF) (besides magnetism and ferroelectricity one sometimes also includes into this class ferroelelasticity, but most often nowadays by MF one means coexistence of magnetism and ferroelectricity). The study of MF materials and of related phenomena was initially started in the former Soviet Union, mainly by the groups of Smolenskii in Leningrad and by Venevtsev in Moscow, and by Schmidt in Switzerland, who coined the very term "multiferroic". After relatively slow start, this field get strong boost at the beginning of 2000-th, starting from works of Ramesh, Kimura and Cheong, and it is now a very well-developed field, with many interesting results and with the promise of important applications [Cheong and Mostovoy, 2007].

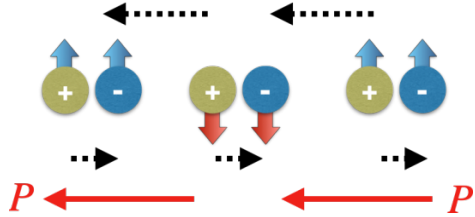

Figure 12. The appearance of electric polarization due to magnetostriction in a system with two types of ions and with magnetic structure up-up-down-down.

As to the physics of MF, there is now pretty good (although of course not complete) understanding of the microscopic mechanisms of MF. All MF can be divided into two big groups. In type-I MF magnetism and ferroelectricity appear independent of one another – although of course connected when both are present! – and are due to different physical mechanisms. Thus e.g. in one of the best MF, $BiFeO_3$ with perovskite structure, ferroelectricity appears first, at ~ 1100K, and is predominantly due to the lone pairs of $Bi^{3+}$. At lower temperatures, below $T_N = 643\ K$, it becomes AFM due to the ordering of spins of $Fe^{3+}$. Similarly, in $RMnO_3$ (R- rare earths like Y, Ho etc.) ferroelectricity and magnetism occur at different temperatures, FE at about 900K, and AFM – at about $T_N \sim 100\ K$. Here FE is connected with the rotation and tilting of $MnO_5$ trigonal bipyramids, and magnetic ordering - with the usual exchange interaction of Mn ions.

Type-II multiferroics, however, are the systems in which the origin of ferroelectricity and magnetism are intrinsically connected and FE appears only in some specific magnetic structures, always in magnetically-ordered state. There are three main microscopic mechanisms of MF behaviour in such systems. One is the magnetostriction: in some types of magnetically-ordered states due to magnetostriction there may appear electric polarization, see e.g. Figure 12. Such are for example $TbMn_2O_5$ or in $Ca_2CoMnO_6$. The second, most widespread and probably the most interesting mechanism of MF in type-II multiferroics is realized e.g. in systems with cycloidal magnetism structures. It was shown in 2005 by Katsura, Nagaosa and Balatsky, and by Mostovoy, that for neighbouring canted spins $\vec{S_i}, \vec{S_j}$ there would appear electric dipole moment, or electric polarization

$$\vec{P} \sim \vec{r}_{ij} \times [\vec{S_i} \times \vec{S_j}], \qquad (4)$$

where $\vec{r}_{ij}$ is the vector connecting these two sites. For cycloidal magnetic structures such dipole moments at every bond add, leading to the net polarization, i.e. to MF behaviours, see Figure 13.

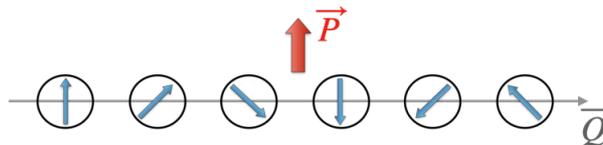

Figure 13. Magnetic cycloid running along x, with spins rotating in the zx plane. According to (4) this results in formation of electric polarization pointing in the z direction.

This is e.g. the mechanism of MF in $TbMnO_3$ - one of the first such systems discovered. Nowadays there are many other systems, mainly TM oxides, with this property.

Microscopically the mechanism of electric polarization in these systems is the inverse DM effect, requiring the presence of relativistic SOC.

There exist also one more, the third microscopic mechanism of MF, also requiring the action of SOC – the SOC-driven modification of *d-p* hybridization. This mechanism can give MF behaviour for example in the helicoidal (proper screw) magnetic structures. But this situation is less common and usually less important in the field of MF.

The study of MF, besides its scientific interest, is largely driven by the possible important applications. The most important of these is the potential possibility to control magnetic memory electrically, without using electric current as in most present magnetic memory devices, but writing and reading magnetic memory just by electric field, like e.g. in gate devices, this avoiding the losses always present when one has electric currents. And one can also make very sensitive magnetic field censors using MF, with the sensitivity approaching that of SQUID but operating at room temperatures. All in all, this field of research is very active nowadays.

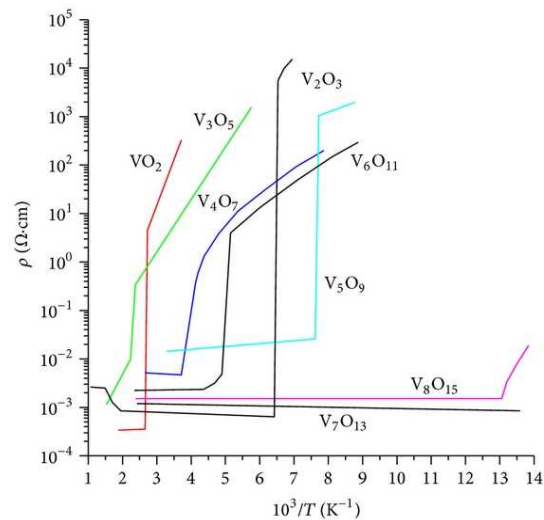

*Figure 14. Metal-insulator transitions in various V oxides as seen from temperature dependence of electrical resistance. Adopted from A.L. Pergament et al., ISRN Condensed Matter Physics 2013, 960627 (2013)*

## 8. Insulator-metal transitions; metallic oxides

As discussed in Section 3, systems with integer number (e.g. one) of electrons per site and with strong correlations, with the Hubbard's on-site interaction $U$ larger than the effective electron hopping $t$ or the respective bandwidths $W = 2zt$, are Mott insulators. When one of these conditions is violated, we can have a transition to a metallic state. One can speak of essentially three specific mechanisms of such insulator-metal transitions (IMT). The first one can be called, following Japanese literature, the bandwidths-controlled IMT [Imada et al., 1998]. If, keeping the number of electrons the same, one increases the electron hopping $t$ (or decreases the Hubbard interaction $U$), one can reach the situation when the bandwidths become larger than the electron repulsion, cf. Figure 4. In this case the system may become metallic, i.e. there will occur in it an insulator-metal (or Mott) transition. This can be reached e.g. by increasing pressure, at which the atoms come closer together and the overlap or electronic functions on neighbouring sites, and corresponding electron hopping, increase[1].

Such transitions can be also driven by temperature. These are the most interesting situations. A good example is presented by the vanadium oxides, see Figure 14. As we see, many of those have a transition from a metallic to an insulating state with change of temperature. Very often such IMT are accompanied by the change of crystal structure, and sometimes – with magnetic ordering. However, they may occur also without change of lattice symmetry. This is seen e.g. in phase diagram of $V_2O_3$ as a function of pressure (or chemical pressure imitating compressed samples) and temperature, Figure 15: the transition line at higher temperatures divides paramagnetic metallic from paramagnetic insulating state, with the same corundum structure (but with the jump in volume at this I-order transition). One sees here also the effect discussed above: $V_2O_3$ becomes metallic at high pressures even at low temperatures. There are many examples of such Mott transitions in different TM oxides.

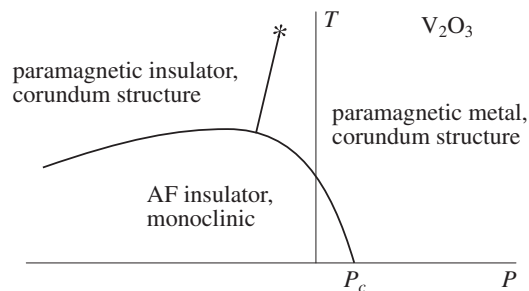

Figure 15. Schematic phase diagram of $V_2O_3$ in pressure (P) and temperature (T) coordinates.

One can also single out the "subtype" of the same mechanism: IMT transition caused not by increase of electron hopping or a bandwidth, but by the reduction of effective Hubbard repulsion $U$. It seems at first glance that it is very difficult to modify it – it is a local property of an ion. But it can change for example if in a many-electron atom or an ion the intra-ionic configuration change, e.g. under pressure. Thus, for example in oxides containing the ion $Fe^{3+}$ (configuration $d^5$, with the spin in the HS state $S=5/2$) such ions transform under pressure to a LS state with the total spin $S=1/2$ (LS state ions have smaller size than the HS ones). One can show that at this transition the effective value of the on-site Coulomb repulsion strongly decreases ($U_{eff}$ contains not only the direct density-density repulsion, but it also depends on the intra-atomic Hund's rule exchange $J_H$, and its contribution strongly changes in going from the HS to LS state). One can call this mechanism the interaction-controlled IMT.

The second situation one meets when one changes the electron concentration, e.g. by doping. By that the system can also become metallic. This is the band-filling-controlled IMT. Important to realise is that the metallic state which can thus be created, still may have narrow bandwidths and strongly-interacting electrons. Therefore, such metallic state can have very different properties from those of the conventional metals like Na or Al. They may strongly deviate from the usual free-electron-like or even from Fermi-liquid behaviour: the temperature dependence of many properties such as resistivity or Hall effect may be different from those of conventional metals; they may display nontrivial magnetic properties partially resembling those of localized electrons, e.g. may develop different types of magnetic ordering (FM, helicoidal etc); they may have different other types of orderings, including charge ordering, discussed above in Section 6. A good illustration may be provided by the rich phase diagram of colossal-magnetoresistance manganites like $La_{1-x}Ca_xMnO_3$, shown in Figure 1. Most interesting, in some such systems one can even have a high-temperature superconductivity, discussed in more details in the next section.

---

[1] This general tendency can be violated in certain specific cases, in particular when in an insulating state TM ions form tightly-bound small clusters, e.g. dimers. Formation of such dimers is usually accompanied by the reduction of the volume if the system, and such small-volume insulating state can be stabilised by pressured. But it is a rather rare exception; as a rule one should expect that with increasing pressure Mott insulators could be transformed into metals.

Yet one more special type of IMT and the corresponding class of metallic oxides can appear in systems in which there exist strong interplay of $d$-electrons of transition metals and $p$-electrons of ligands, e.g., oxygens. As discussed in Section 3, one has in general to include oxygen $p$-electrons in the discussion, and this can become essential if oxygen $p$-levels lie close to $d$-levels of TM. One can characterize it by the CT energy, needed to transfer an electron from $O^{2-}$ ion to $d$-levels of TM. As mentioned in Section 3, if $\Delta_{CT}$ is very large one can exclude $p$-electrons and go to the effective description including only $d$-electrons, with some effective parameters. But in some cases, especially in systems with high valence of TM, $\Delta_{CT}$ can be small or even negative. In such situation it may be energetically favorable to transfer some $p$-electrons from oxygens into $d$-shells of TM, thus creating oxygen holes $\underline{L}$ - oxygen ions with $p^5$ configuration). Oxygen hole thus created can start to move in a crystal, creating metallic state. One can call this situation self-doping: one changes concentration of $d$-electrons and $p$-electrons not by external doping as in band-filling-controlled IMT, but such process occurs spontaneously due to redistribution of electrons between $p$-shells of oxygens and $d$-shells of TM. The properties of such states can be very unusual. Such states can be very important in many materials, in particular in the cuprate High-Tc superconductors discussed in the next section.

## 9. Superconductivity in oxides

Usually when speaking on superconductors one has in mind good metals like Pb or Al. Superconducting pairing in those is ascribed to electron-phonon interaction. Magnetism was always considered as the greatest enemy of superconductivity. And in most cases it is indeed true. All the more surprising was the discovery of High-Tc superconductivity by Bednorz and Mueller in 1987 in completely different class of materials: in cuprates, TM oxides usually considered good for magnetism but not for superconductivity. This discovery caused enormous activity continuing to this day and attracted common attention to the physics of correlated electron systems, to TM oxides in general and to cuprates in particular [Keimer et al., 2015].

Cuprate High-Tc superconductors, such as $La_{2-x}Sr_xCuO_4$ (LSCO) or $YBa_2Cu_3O_7$ (YBCO) contain as the main building blocks $CuO_2$ planes shown in Figure 16, i.e. they can be considered as "2D perovskites".

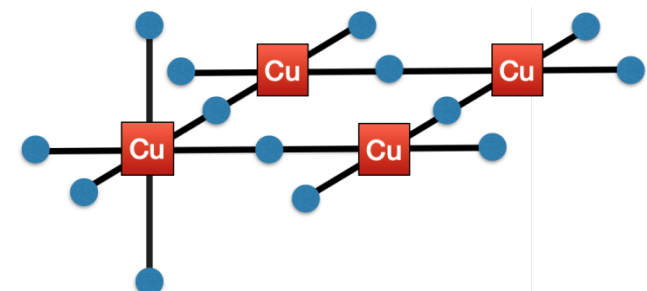

*Figure 16 Basic structure of CuO2 layer constituting the main building block of High-Tc cuprate superconductors. Blue balls are oxygen ions (one or two apical oxygens can be missing in some members of this family)*

Formally the valence of Cu in them is $Cu^{2+}$ in undoped LSCO or in oxygen-depleted YBCO, which are actually Mott (or rather charge-transfer) insulators with magnetic ordering. $Cu^{2+}$ is a very strong JT ion, and in these systems it is typically located either in very strongly elongated octahedra, or is five-fold or even four-fold coordinated, i.e. located in square pyramids or just in oxygen squares. The hole in $d^9$ configuration of $Cu^{2+}$ is sitting on the upper $x^2$-$y^2$ orbital, rather well separated from the rest of $d$-states, cf. the right part of Figure 2a. Superconductivity appears in these systems when they are hole-doped, so that the formal valence of Cu becomes e.g. $Cu^{2+x}$ in $La_{2-x}Sr_xCuO_4$. But "$Cu^{3+}$" thus created is known to be a negative charge-transfer ions, see Section 8 above, i.e. the large fraction of thus created holes are actually in oxygens. The resulting state can be represented as $Cu^{2+}\underline{L}$, with the spins of $Cu^{2+}$ and of the, predominantly oxygen, holes forming singlet states - the Zhang-Rice singlets. In effect one can often reduce the description of cuprates to the one-band Hubbard model, with undoped magnetic systems corresponding to the half-filled case thereof, the holes introduced by doping - Zhang-Rice singlets – playing the role of holes in such nondegenerate Hubbard model.

But in fact, besides the magnetic and superconducting states, one has found in high-Tc cuprates much more complicated behaviour – the phases of "strange metals" with non-Fermi-liquid behaviour, the phase with a pseudogap, with different presumed types of ordering - see the schematic phase diagram in Figure 17.

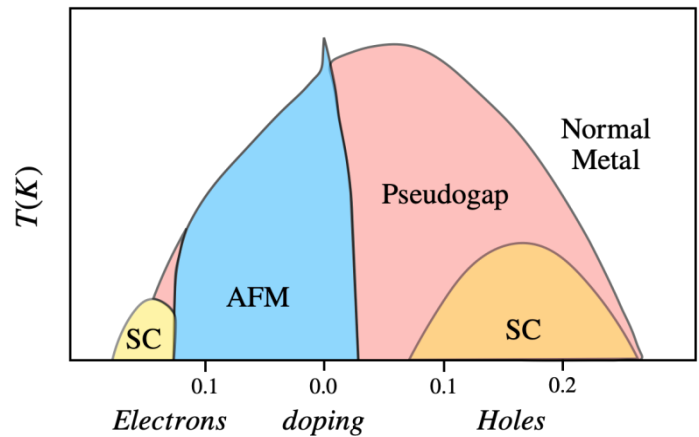

*Figure 17. Phase diagram of superconducting cuprates; superconducting region is shown in blue. SC stands for superconducting, AFM – antiferromagnetic phases.*

Despite enormous progress in this field, some basic questions, such as the very mechanism of High-Tc superconductivity, and many properties of the normal state for these compositions, still remains open. It is established that in contrast to the usual superconductors the superconducting pairing in cuprates, described by the pairing wave function or the order parameter $\Delta(\vec{k})$, is singlet but not of the $s$-wave type as in most usual superconductors, but is of the $d$-wave type, $\Delta(\vec{k}) \sim |\Delta|(\cos k_x - \cos k_y)$, i.e. it has nodes in certain directions in $k$-space. As to the mechanism of High-Tc superconductivity, it is accepted that it is predominantly not due to the usual electron-phonon interaction, but is largely due to interelectronic interactions. The proximity of High-Tc region to magnetic states is apparently not accidental, magnetic effects seems to play crucial role in establishing superconducting pairing. But the details of this process are still a matter of hot debates, 30 years after the discovery of this phenomenon.

Besides cuprates, yet one more class of high-Tc superconductors was discovered later on, also containing strongly corrected electrons – the iron-based superconductors like LaOFeAs or $\beta-$FeSe. Apparently, many properties of these have much in common with those of cuprates, although there are also important differences. As these materials do not belong to oxides, we will not discuss them further.

**10. Various useful properties and applications of magnetic oxides**

When speaking of applications of TM oxides, one can point out several specific directions. The first one is the use of their magnetic properties per se. It was in fact the first know in history application – the use in ancient China of magnetite $Fe_3O_4$, which is already magnetic at room temperature, for orientation of buildings under construction and for navigation: the first ever compass, on the basis of magnetite, was invented in China about 2000 years ago. Magnetic properties of oxides were used recently in such applications as magnetic memory in computers: the early memory on ferrite coils, more recently using $Fe_2O_3$ in older hard discs.

More elaborate applications rely on the dependence of various properties of TM oxides such as e.g. electrical resistivity on magnetic order. Thus there are attempts to use colossal magnetoresistance in manganites and some other materials as sensors. Recently there was a big activity in studying topological effects in particular in systems with strongly correlated electrons, among them in oxides. Thus for example topological Hall effect was observed in $SrRuO_3$ films.

There are the attempts to use magnetic oxides for elaborate thermopower applications. The idea is to use the extra entropy often present in these systems – magnetic, orbital entropy – for entropy transfer, which could make these systems promising candidates for new class of thermoelectric materials.

Strong magnetoelastic coupling and magnetostriction, typical for TM system with unquenched orbital moments, e.g., containing such ions as $Fe^{2+}$, $Co^{2+}$, can be used for non-destructive control of metallic products such as gas and oil pipes, etc.

For practical purposes, especially in electronics, the use of thin films is of paramount importance. Investigation of films and multilayers of different materials is going on very rapidly at the moment. TM oxides present an important playground in this direction. Three main factors can lead to the modification of the behaviour of films as compared to the bulk. The first one, common for all films independent on their particular type, is spatial quantization, caused by the finite thickness of films restricting motion of electrons, magnons etc in perpendiculars direction. Two other factors are more specific for TM oxides. First, due to strain imposed by the substrate, both magnetic and especially orbital structure of the material can change. For example, orbital occupation at the first few layers might be very different from that in the bulk. This in its turn can lead to change of exchange interaction and to different type of magnetic ordering in thin films as compared to the bulk. The third factor, important for TM films and multilayers, is the possible charge redistribution connected with the possibility to change the valence of TM ions. Thus, for example when one makes contact of $LaMnO_3$ with $LaNiO_3$, with the valence in the bulk being 3+ for both TM ions, on the interface it is favorable to transfer electrons from Mn to Ni, forming on the contact $Mn^{4+}$ and $Ni^{2+}$. Indeed, the trivalent Ni is an ion with very small or even negative charge transfer gap and is not very stable: energetically $Ni^{2+}$ is much more favorable. On the other hand, to transform $Mn^{3+}$ to $Mn^{4+}$ does not cost much energy. In effect this charge and valence redistribution on the contact indeed takes place in this situation, and similar phenomena should be considered also for some other combination of TM oxides in a contact.

Some TM oxides can serve as very useful substrates. The most important is $SrTiO_3$, but also many other TM oxides are used for this purpose. One can also use by that their transport and magnetic properties. Thus, $SrRuO_3$ can serve as a ferromagnetic metallic substrate, which is necessary for some applications. All in all, the behaviour of TM oxides in thin films and multilayers may be quite diverse, such systems display very rich and interesting properties and can promise important applications.

One more big field in which magnetic oxides are also used widely, is spintronics – the field based on the idea to use not charge but spin of an electron for many applications, e.g. for memory and for logical elements in quantum computers, for controlling transport properties etc. This big and very active field is discussed in other articles in this volume, thus we will not dwell on this anymore; suffice is to say that it is just the diversity of the properties of TM oxides which makes them very useful in different ways in spintronics devices.

And one more important point: one should not forget that the TM oxides are quite widespread in nature and play an important role in many natural phenomena. Oxygen gives almost 50% of the total mass of Earth's crust, ocean water and atmosphere. Also, iron is quite abundant (~5%). And this iron mostly exists in Earth's crust in form of oxides which are essentially magnetic: magnetite $Fe_3O_4$ – ferrimagnetic, hematite $Fe_2O_3$ (the main component of ordinary rust) - AFM with eventual spin canting. Similarly, other TM and RE metal oxides are well represented in the Earth's crust. They are often the main ores used for metal production, especially for iron.

Most of TM oxides in the Earth's crust have some magnetic properties, although for the laymen they are not "magnetic" - they do not attract small metallic objects such as needles or nails: most of these are AFM of some kind. But their magnetic properties are widely used in geophysics for the search of novel useful mineral deposits; magnetometry is one of the most powerful tools in modern archeology, etc. And minerals are inexhaustible source of novel magnetic materials with often exotic properties. Thus, most of novel kagomé systems originate from minerals, which one sees from their names: herbertsmithite, volborthite, kapellasite, atacamite, to name just a few.

Even when a mineral itself is nonmagnetic, adding TM impurities may drastically modify its properties. Thus, the color of many precious and semi-precious stones like ruby or amethyst are due to TM impurities. And such impurities are the main ingredients in modern laser materials, starting from the ruby lasers.

And the last point: magnetic oxides are quite important for many living creatures, they were already in use by them for ages. Long before the compass with the use of magnetite was invented in ancient China, the same principle and the same material was used by birds for the orientation, and there are many other similar examples. Even bacteria may be sensitive to magnetic field, e.g., magnetotactic bacteria, mostly containing microcrystals of the same magnetite $Fe_3O_4$. Thus, summarizing this article, we may conclude that magnetic oxides are extremely versatile, interesting and important materials, as for their very rich and diverse properties, for their already existing and future applications, but also for many natural phenomena.

**Summary.** Thus, summarizing this article, we may conclude that magnetic oxides are extremely versatile, interesting and important materials, as for their very rich and diverse properties, for their already existing and future applications, but also for many natural phenomena. In this chapter we describe the main physical phenomena in transition metal oxides determining their properties – mostly magnetic, but also other interesting phenomena, such as insulator-metal transitions, magnetoelectricity and multiferroicity, and even High-Tc superconductivity in some of them. Basic theoretical concepts are shortly described in this chapter, and diverse properties of magnetic oxides are illustrated on many examples. We conclude by shortly mentioning several applications of magnetic oxides.

## Acknowledgments

D. Kh. was supported by DFG through grant 277146847-CRC-1238, while S.S. via project Quantum 122021000038-7.

## Abbreviations

AFM – antiferromagnetic, antiferromagnetism
DM - Dzyaloshinskii-Moriya
CF – crystal-field
CT – charge-transfer
IMT - insulator-metal transitions
JT – Jahn-Teller
High-Tc - high-temperature
FE - ferroelectricity
FM – ferromagnetic, ferromagnetism
HS – high-spin
LS – low-spin
ME - magnetoelectric effect
MF - multiferroic
RE – rare-earths
RKKY - Ruderman-Kittel-Kasuya-Yosida
SOC – spin-orbit coupling
TM - transition metals